\documentclass[prd,twocolumn,aps,showpacs,nofootinbib,nobibnotes,superscriptaddress]{revtex4}

\usepackage{epsfig}
\usepackage{amssymb}

\newcommand{\de}{\delta}
\newcommand{\ga}{\gamma}
\newcommand{\la}{\lambda}
\newcommand{\om}{\omega}
\def\lsim{\, \raise 0.4ex\hbox{$<$}\kern -0.8em\lower 0.62
ex\hbox{$\sim$} \,}
\def\gsim{\, \raise 0.4ex\hbox{$>$}\kern -0.8em\lower 0.62
ex\hbox{$\sim$} \,}
\begin{document}


\title{Testing Lorentz Invariance Violation with the WMAP Five Year Data}


\author{Tina Kahniashvili}
\email{tinatin@phys.ksu.edu} \affiliation{Department of Physics,
Kansas State University, 116 Cardwell Hall, Manhattan, KS 66506,
USA}\affiliation{Department of Physics, Laurentian University,
Ramsey Lake Road, Sudbury, ON P3E 2C6, Canada}  \affiliation{E.
Kharadze  Abastumani Astrophysical Observatory, Ilia Chavchavadze
State University, 2A Kazbegi Ave, GE-0160 Tbilisi, Georgia}

\author{Ruth Durrer}
\email{Ruth.Durrer@physics.unige.ch} \affiliation{D\'epartement
de Physique Th\'eorique, Universit\'e de Gen\'eve, 24 quai Ernest
Ansermet, 1211 Gen\'eve 4, Switzerland}

\author{Yurii Maravin}
\email{maravin@phys.ksu.edu} \affiliation{Department of Physics,
Kansas State University, 116 Cardwell Hall, Manhattan, KS 66506,
USA}

\begin{abstract}
We consider different renormalizable models of Lorentz invariance
violation. We show that the limits  on birefringence of the
propagation of  cosmic microwave background photons from the five
year data of the Wilkinson microwave anisotropy probe (WMAP) can
be translated into a limit of Lorentz symmetry violation. The
obtained limits on  Lorentz invariance violation are stronger
than other published limits. We also cast them in terms of limits
on a birefringent effective photo ``mass'' and on a polarization
dependence of the speed of light.

\end{abstract}


\maketitle
\section{Introduction}

The principal spacetime symmetry of particle interactions in the
standard model is Lorentz invariance. Experiments confirm Lorentz
symmetry at all currently accessible energy scales of up to 2
TeV. This scale will be extended  shortly to 14 TeV with the
Large Hadron Collider (LHC) at CERN. Although present experiments
confirm Lorenz invariance to a good precision, it can be broken
in the very early Universe when energies approach the Planck
scale. There are a number of extensions of the standard model of
particle physics and cosmology that violate Lorentz invariance
(for reviews see Refs.~\cite{k05b,shore,m05,jlm05,recent}).

As it can be expected, Lorentz invariance violation  (LV) affects
photon propagation (the dispersion relation), and generically
results in a rotation of linear polarization (birefringence).
Other effects include new particle interactions such as a photon
decay and vacuum Cherenkov radiation~\cite{jlm05}. All these
effects can be used to probe Lorentz invariance.  The dispersion
measure  test is based on a phenomenological energy dependence of
the photon velocity~\cite{a98} (see also Refs.~\cite{sarkar} for
reviews and Refs.~\cite{jlm03,bwhc04,MP06} for recent studies of
this effect; early discussions include Refs.~\cite{16'};
Refs.~\cite{a98,bwhc04,MP06} consider Lorentz symmetry violating
models which preserve rotational and translational invariance but
break boost invariance).

Several models of LV predict frequency  dependent effects. Such
high energy Lorentz invariance breaking are discussed in Refs.
\cite{km01,at01,myers}. Refs.~\cite{GLP05} study generalizations
of electromagnetism, motivated by this kind of Lorentz invariance
violation. On the other hand,  LV associated with a Chern-Simons
interaction \cite{rj98,cfj90} affects the entire spectrum of
electromagnetic  radiation, not just the high frequency part, and
induces a frequency independent rotation of polarization (see
Sec.\ 4 of Ref.~\cite{shore} and Sec.~III of this work).

To determine the effects induced by Lorentz symmetry violation, it
is useful to consider the analogy with the propagation of
electromagnetic waves in a magnetized plasma as outlined in
Refs.~\cite{cfj90,km01,jlm03,mmu05,kgr07}. Using the well known
formalism for the propagation of light in a magnetized plasma, is
easy to see that for Lorentz symmetry violating models which
depend also on polarization and not only on frequency, the
rotation measure  constrains the symmetry breaking scale more
tightly than the dispersion measure, see
Refs.~\cite{GLP05,myers,kgr07}.

The propagation of ultra-high energy photons represents a
promising possibility to probe Lorentz symmetry~\cite{sigl08}.
Gamma Ray Bursts (GRB) are astrophysical objects located at
cosmological distances which emit very energetic
photons~\cite{a98}; reviews describing cosmological tests
involving GRBs are e.g. Refs.~\cite{piran,m05}, for recent studies
see \cite{Albert:2007qk}. After the observation of highly
linearly-polarized $\gamma$-rays from GRB021206 has been
reported~\cite{polarimetry}, Refs.~\cite{mitrofanov,jlms04} have
proposed to  test Lorentz symmetry violation with the rotation
measure by analysis of GRB polarization. Even though this
measurement has been strongly contested~\cite{pol2}, there is
evidence that the $\gamma$-ray flux from  GRB930131 and GRB960924
is consistent with more than $35\%$ and $50\%$  polarization,
respectively~\cite{Willis}. However, the issue of  polarization
of GRB $\gamma$-rays is still under debate and additional $X$-ray
studies are needed to either confirm or disprove polarization of
$\gamma$-rays~ from GRB's~\cite{25}.

 In this paper
we mainly consider renormalizable models of LV as described in
Ref.~\cite{shore}. We use the very well understood and measured
temperature anisotropy and polarization of the cosmic microwave
background (CMB) to constrain Lorentz symmetry violation. These
data have been proposed as a probe of Lorentz invariance in the
Universe in Refs.~\cite{Lue,Feng,mewes,kam}. In our study we use
the WMAP 5 year limits on birefringence~\cite{WMAP} and obtain
limits which are significantly more stringent than those obtained
from radio galaxy polarimetry~\cite{cfj90}.

As we shall see below, generically Lorentz symmetry violation
leads to birefringence, i.e. a photon dispersion relation which
depends on polarization. This leads to a rotation of the CMB
polarization which induces parity-odd cross correlations, such as
Temperature-$B$-polarization and $E$-$B$-polarization \cite{Lue}.
These correlators vanish in models which preserve parity.
Generally speaking, the effect is similar to that induced by a
homogeneous magnetic field~\cite{Bfield,Bconst}. In this paper we
use the WMAP-5 year  limit on the rotation measure~\cite{WMAP} to
contrain Lorentz invariance violating theories.

\section{Lorentz invariance violation: general description}

For methodological purpose let us first briefly
summarize the usual Faraday rotation effect. We consider an
electromagnetic wave with frequency $\omega$ and spatial wave
vector ${\bf k}$, $k\equiv |{\bf k}|$ propagating in a magnetized
plasma. A linearly polarized wave can be expressed  as
superposition of left ($-$) and right ($+$) circularly polarized
waves. In a magnetized plasma, a homogeneous magnetic field
induces a difference in the phase velocity of left and right
handed waves. This causes a rotation of the polarization, called
Faraday rotation~\cite{krall}. The group velocity of the wave also
differs from $c$. These two effects can be expressed in terms of
the refractive indices defined by $k_{\pm} =n_{\pm}\omega$ where
$k_{\pm}$ denotes the wave number for right and  left handed
 waves. The indices $ n_{{\pm}}$ are~\cite{krall}
\begin{equation}
n_{\pm}^2  = 1-\frac{\om_p^2}{\om(\om\pm\om_c)} \simeq
1-\frac{\om_p^2}{\om^2} \pm \frac{\om_p^2\om_c}{\om^3}  ~,
\label{eq:05}
\end{equation}
Here $\om_p = 4\pi e^2n_e/m_e$ is the plasma frequency and $\om_c
=eB/m_e$ is the electron cyclotron frequency for the magnetic
field $B$ (see Sec. 4.9 of Ref.~\cite{krall}).

The magnitude of both the dispersion measure, due to the
different group velocities, and the rotation measure, i.e., the
rotation of polarization, are proportional to the photon travel
distance $\Delta l$,
 \begin{eqnarray}
\Delta t_{\pm} &=& {\Delta l} \left( 1 - \frac{\partial
k_{\pm}}{\partial \omega} \right), \label{time-delay}
\\
\Delta \alpha &=& \frac{1}{2}(k_{+} - k_{-}) \Delta l.
\label{angle}
\end{eqnarray}
Here, $\Delta t_{\pm}$ is the difference between  the  travel
time of a right-handed (left-handed) photon and that of a photon
traveling at the speed of light, and $\Delta \alpha$ is the
rotation of the angle of  polarization.

Faraday rotation is widely used in astrophysics to measure
magnetic fields in galaxies and clusters (see Ref.~\cite{Valee}
for a review and references therein). In cosmology, Faraday
rotation of CMB photons~\cite{Bfield,faraday} has been used to
constrain the amplitude of a homogeneous as well as a stochastic
cosmological magnetic field~\cite{fine}.

In the following, we show that Lorentz symmetry violation leads to
a modification of Maxwell's equations~\cite{myers,GLP05}
analogous to the modifications described above.

Following Ref.~\cite{shore}, the most general  renormalizable
form of Lorentz symmetry violation can be expressed by two
additional terms in the action (we set $\hbar=c=1$)
\begin{equation}
\Gamma_{\rm LV} = \int d^4 x\sqrt{-g} \left[ K_{\mu\nu\lambda\rho}
F^{\mu\nu}F^{\lambda\rho} - \frac{1}{4} L^\mu A^\nu \tilde
F_{\mu\nu}\right], \label{Gamma}
\end{equation}
where Greek indices ($\mu,\nu,\lambda,\rho$) denote time-space
coordinates,  $F^{\mu\nu}$ is the electromagnetic field strength
tensor, $\tilde F_{\mu\nu}
={\epsilon_{\mu\nu}}^{\lambda\rho}F_{\la\rho}$ is its dual,
$\epsilon_{\mu\nu\lambda\rho}$ is the totally antisymmetric
tensor normalized such that $\epsilon_{0123} = \sqrt{-g}$ and
$A^\nu$ is the vector potential. The four-vector
 $(L_\mu) = (L_0, {\bf L})$
has the dimension of mass and describes a super-renormalizable
(dimension 3) coupling and $K_{\mu\nu\lambda\rho}$ is a
renormalizable, dimensionless coupling giving raise to a
dimension 4 operator. We want to break Lorentz symmetry, but keep
conformal invariance of electrodynamics in this work. For this we
have to ask that the components of ${K_{\mu\nu}}^{\la\rho}$ and
$L_\mu$ be independent of conformal transformations of the
metric. In the cosmological setup with $g_{\mu\nu}
=a^2(t)\eta_{\mu\nu}$, the above action is then independent of
the scale factor $a(t)$. I.e. in a conformally flat spacetime the
action is like in flat space. To see this note that the forms
$A_\mu$ and $F_{\mu\nu}$ are independent of the metric hence
$K_{\mu\nu\lambda\rho}F^{\mu\nu}F^{\lambda\rho}$ and $L^\mu A^\nu
\tilde F_{\mu\nu}$ scale like $a^{-4}$ which is canceled by
$\sqrt{-g}=a^4$.

The tensor $K_{\mu\nu\lambda\rho}$ has the same symmetries as the
Riemann tensor and we only consider its trace-free part which is
analog to the Weyl tensor (the trace part also leads to
dispersion measure but not to birefringence, we therefore do not
consider it here). Even though we apply the formalism used for
the Weyl tensor below, we do not consider $K_{\mu\nu\lambda\rho}$
to be the Weyl tensor which of course vanishes in a (unperturbed)
Friedmann universe. The most plausible origin for the Lorentz
violating terms in (\ref{Gamma}) is that $K_{\mu\nu\lambda\rho}$
or the vector $L_\mu$ stem from the non-vanishing vacuum
expectation value of some dynamical field and the action
(\ref{Gamma}) therefore represents a spontaneous rather than
explicit breaking of Lorentz symmetry. However, for the following
discussion the origin of the Lorentz violating terms is not
relevant.

Both terms in Eq.~(\ref{Gamma}) lead to birefringence but the
frequency dependence is different. The first term in the action
$\Gamma_{\rm LV} $ can be computed within the Newman-Penrose
formalism, which is usually applied for the Weyl
tensor~\cite{shore}. We consider a plane wave with conformal wave
vector $(k_{\mu})=(\om,{\bf k})$. We normalize the scale factor
to unity today, $a_0=a(t_0)=1$, so that conformal frequencies or
length scales correspond to physical scales today. In terms of
the conformal wave-vector, the dispersion relation is like in
flat space where it has been derived in Ref.~\cite{shore},
\begin{equation}
\omega^2 = k^2  \mp {8} \omega^2 |\Psi_0| \,. \label{term1}
\end{equation}
Here $\Psi_0$ is the analog of the Newman-Penrose scalar (for
more details see \cite{shore}),
$$
\Psi_0 =-a^{-4}\left[K_{0i0j} - K_{0ilj} n^l  - K_{kilj}n^k n^l
\right] m^im^j\,,
$$
where ${\bf m}$ and $\bar{\bf m}$ represent the left and right
circular polarization basis vectors and ${\bf n}={\bf k}/k$ is
the photon propagation direction. We normalize ${\bf n}$ and
${\bf m}$ with the flat metric, $n^in^j\de_{ij}=m^i\bar
m^j\de_{ij}=1$, and multiply the expression with the correct
power of the scale factor, $a^{-4}$, so that, given the scaling
of the tensor $K$, one sees explicitly that $\Psi_0$ is
independent of the scale factor. (Latin indices indicate spatial
components of a vector or tensor.)

The second term in the action $\Gamma_{\rm LV}$ leads to the
dispersion relation~\cite{cfj90,shore}
\begin{equation}
(k_\mu k^\mu)^2 + (k_\mu k^\mu) (L_\nu L^\nu) - (L_\mu k^\mu)^2
=0, \label{second}
\end{equation}
To first order in the small parameters $L_0$ and
$\sqrt{\de^{ij}L_iL_j}\equiv L$ one has
\begin{equation}
\omega^2 = k^2 \mp \om(L_0 - L\cos\phi),
 \label{second1}
\end{equation}
where $\phi$ is an angle between the photon propagation direction
and the vector ${\bf L}$, $\cos\phi=({\bf L}\cdot{\bf n})/L$.
Note the similarity of the expressions~(\ref{term1}) and
(\ref{second1}) with the corresponding ones following from
Eq.~(\ref{eq:05}).

To be as general as possible, we rewrite the dispersion relation
for both types of Lorentz symmetry violation in the form (see also
\cite{kgr07}),
\begin{equation}
k^2 = \omega^2 \left[ 1\pm \left(\frac{M}{M_{\rm PL}}\right)
\left(\frac{\omega}{M_{\rm PL}}\right)^{N-4} \right],
\label{dispersion-gen}\end{equation} where $M_{\rm PL}$ is the
Planck mass, $M_{\rm PL} \simeq 1.2 \times 10^{19}$ GeV, $N$ is
the dimension of the Lorentz symmetry violating operator and $M$
is a mass scale of the model. For $N=4$, the birefringent part is
independent of the photon energy and we have $8\Psi_0=M/M_{\rm
PL}$. For $N=3$ the Planck mass cancels out and the birefringent
term is inversely proportional to the photon energy. The mass
scale is $M=L_0-L\cos\phi$. Generally speaking, the smaller $M$,
the weaker LV. For $N=4$, LV is frequency independent and the
amplitude of the effect is of order $M/M_{\rm Pl}$, while for the
super-renormalizable case, $N=3$, LV is strongest at low
frequencies, $\om<M$.
  Our aim is to limit the function
$$\gamma(\om)
\equiv \left(\frac{M}{M_{\rm PL}}\right) \left(\frac{\om}{M_{\rm
PL}}\right)^{N-4}$$ from CMB birefringence. This ansatz can also
be applied to non-renormalizable models with higher dimension
operators. For $N\ge 5$, $M\neq 0$ indicates that there is LV at
frequencies $\om \gsim M_{\rm Pl}(M_{\rm Pl}/M)^{\frac{1}{N-4}}$.
\vspace{5pt}

\section{Results}

To compute the CMB polarization rotation angle induced by Lorentz
symmetry violation, we follow the analogy with photon propagation
in a magnetized medium which yields $n_{\pm} = 1 \pm
\gamma(\om)/2$. Using Eq.~(\ref{angle}), we obtain
\begin{equation}
\Delta \alpha^{(LV)} = \frac{1}{2} \omega \gamma(\om) \Delta l
\label{angle1}.
\end{equation}
In the case $N=4$, $\gamma$ is frequency independent, hence
$\Delta \alpha^{(LV)}$ grows linearly with frequency. In this
case, and for all models with higher dimension operators, the
best limits can in principle be obtained from high frequency
photons (for example GRB $\gamma$-rays \cite{mitrofanov,jlms04}),
while CMB photons are less affected. However, the fact that the
theory of CMB anisotropies and polarization yields that both $TB$
and $EB$ polarization have to vanish in standard cosmology, while
the polarization of GRB's is still under debate, at present, a
test using CMB data is to be preferred. Another advantage is that
for the CMB the distance $\Delta l \simeq H_0^{-1}$ is maximal.

In the dimension 3 model, $\Delta \alpha^{(LV)} = -\frac{1}{2}
(L_0-L\cos\phi) \Delta l$, is frequency-independent. In
Ref.~\cite{cfj90} the above result is applied to polarization
data from distant radio galaxies, $\Delta \alpha < 6^{\rm o}$ at
$95\%$ C.L. at redshift $z \sim 0.4$.
 The constraint obtained if
Ref.~\cite{cfj90} is $|L_0 - L\cos\phi | \leq 1.7 \times 10^{-42}
h_0$ GeV, where $h_0\simeq 0.7$ is the present Hubble parameter
in units of $100$ km ${\rm s}^{-1}$ ${\rm Mpc}^{-1}$.

We use the recent WMAP-5 year constraints on the rotation angle
of the CMB polarization plane (combined constraints from the low
and high multipole number, $l$,  data assuming a constant $\Delta
\alpha $ across the entire multipole range), $-5.9^{\rm o} <
\Delta \alpha < 2.4^{\rm o}$ at 95\% C.L. and $\Delta \alpha =
-1.7^{\rm o} \pm 2.1^{\rm o}$ at 68\% C.L.~\cite{WMAP} (Sec.
4.3).  Assuming Gaussian errors, it is straightforward to convert
this to the following limits on the absolute value of rotation
angle,
\begin{eqnarray}
|\Delta \alpha |_{\rm obs} &\leq & 4.90^{\rm o}~~~~~~~{\rm at}
~~~~95\%~~{\rm C.L.}~, \label{95}\\| \Delta \alpha |_{\rm obs}
&\leq& 2.52^{\rm o}~~~~~~~{\rm at}~~~~68\%~~{\rm C.L.}~.
\label{68}
\end{eqnarray}
We adopt $\Delta l \simeq 9.8 \times 10^9 h_0^{-1}$ years. We
express our results in terms of $\nu_{100}=\nu/100{\rm GHz}$ to
keep them as independent of the CMB band frequency as possible.

Using Eq.~(\ref{angle1}), we find the following limit on the
function $\gamma(\nu)$ with $\om=2\pi\nu$:
\begin{eqnarray}\label{e:gamma}
\gamma(\nu) &\leq& 8.6 \times 10^{-31}
\nu^{-1}_{ 100}h_0 ~ \mbox{ at 95\% C.L.,}\\
\gamma(\nu) &\leq& 4.4 \times 10^{-31} \nu^{-1}_{ 100}h_0 ~
\mbox{ at 68\% C.L.,}~.
\end{eqnarray}
We can also express the limit on $\gamma$ in terms of a limit for
the mass scale $M$ or the dimensionless parameter $M/M_{\rm Pl}$
\begin{equation}
\frac{M}{M_{\rm Pl}} \lsim 8.6\times10^{-31}
 \left(3\times 10^{31}\right)^{(N-4)}\nu_{100}^{3-N}h_0  ~
          \mbox{ at 95\% C.L.}~.
\end{equation}
For $N>4$, these limits are not very interesting, while for $N=4$
or $N=3$ 'naturally expected' values of the parameters are ruled
out. More precisely, for the models considered we constrain the
dimensionless
 scalar $\Psi_0$ for the $N=4$ case,
$$|\Psi_0| \leq   1.1 \times 10^{-31} h_0 \nu^{-1}_{100}  ~ \mbox{  at
 95\% C.L.,} $$ while we find for $N=3$
$$|L_0-L\cos\phi| \leq 3.6 \times 10^{-43} h_0\mbox{GeV \qquad
at 95\% C.L.}~. $$ This is almost an order of magnitude better
than the limit obtained in Ref.~\cite{cfj90}.

We can also introduce an effective photon ``mass'' by writing the
modified dispersion relation in the form $\omega^2 = k^2 \pm
m_\gamma^2$ with
$$m_\gamma^2 = \omega^2 \gamma(\om) = M \omega
\left(\frac{\omega}{M_{\rm PL}}\right)^{N-3} =
2\frac{\Delta\alpha}{\Delta l}\omega\,. $$ For $N> 2$  this is
not a mass in the usual sense of the energy of the particle at
rest, but rather a measure for the modification of the dispersion
relation which tends to zero with frequency. For the
renormalizable dimension 4 and 3 operators considered in this work
we have $ m_{\gamma}^{(4D)}(\om) = 2 \omega |2\Psi_0|^{1/2}$ and
$m_\gamma^{(3D)} = \left[\omega(|L_0 - L\cos\phi|)\right]^{1/2}$
respectively. As in  Ref. \cite{mitrofanov} we can interpret our
result also in terms of a polarization dependent group velocity,
\begin{equation}
 v_{\pm} = 1\pm \frac{N-2}{2}\frac{M}{M_{\rm Pl}}
\left(\frac{\omega}{M_{\rm Pl}}\right)^{N-4} = 1\pm
    \frac{N-2}{2}\ga(\om)~.
\end{equation}
Ref.~\cite{mitrofanov} only studied the cases $N\ge 5$. From
Eq.~(\ref{e:gamma}) we derive the constraint on the effective
birefringent mass,
\begin{equation}\label{e:mass}
m_\gamma \leq 3.8 \times 10^{-19}\left( h_0
\nu_{100}\right)^{1/2} {\rm eV~~ at}~ 95\%~{\rm C.L.}
\end{equation}
Note that left and right handed photons have effective square
masses of opposite sign. For the velocity difference this implies
\begin{equation}\label{e:vel}
|v_+-v_-| \leq \left\{\begin{array}{lll}
  8.6\times 10^{-31} h_0v_{100}^{-1} & \mbox{  at 95\% C.L.,} & \mbox{for } N=3\\
  1.7\times 10^{-30} h_0v_{100}^{-1} & \mbox{  at 95\% C.L.,} & \mbox{for } N=4~.
\end{array} \right.
\end{equation}
The limits on $m_\gamma$ are model independent because $m_\gamma$
only depends on the directly measured rotation angle
$\Delta\alpha$ and on the frequency.

If $L \ll L_0$, we can safely neglect the angular dependence, and
assume that $m_\gamma^{(3D)} = \sqrt{\omega L_0}$. However, if $L
\gg L_0$, the modification of the photon dispersion becomes
direction dependent, and must be averaged over all sky for the
CMB photons. Then, the rotation angle can be estimated by the
two-point correlation function, $i.e.$, $\Delta \alpha_{\rm eff}
= \langle |\Delta \alpha|^2\rangle^{1/2}$. A rough estimate leads
to a pre-factor $\sim1/\sqrt{2}$. In a more detailed analysis the
presence of ${\bf L}$  breaks rotational symmetry and leads to
off-diagonal correlations in the temperature anisotropy and
polarization spectra analog to the effects on the CMB by a
constant magnetic field~\cite{Bconst,faraday}. To take this fully
into account requires to estimate the CMB Temperature-B
polarization, E- and B-polarization cross correlations, as well
as B-polarization spectra due to the Lorentz symmetry violating
vector field ${\bf L}$, and to compare theoretical estimates with
the corresponding CMB anisotropy and polarization data. Also the
scalar $|\Psi_0|$ of the 4D model breaks rotational symmetry and
taking the direction dependence of $\Delta\alpha$ into account is
relatively complicated. This breakdown of statistical isotropy
can also be tested using the bipolar power spectrum introduced in
Ref.~\cite{soura}.

We shall address this issue in future work, but even though the
limits may improve somewhat, we do not expect them to change
significantly.
$$
$$

\section{Conclusions}

The obtained bound on a birefringent effective photon
 mass is below the limit for a standard
photon mass given by the particle data group~\cite{particle},
$m_\gamma \leq 3\times 10^{-19}< 10^{-18}$ eV, but is less
stringent than the limit from galactic magnetic fields which is,
however model dependent~\cite{dvali}. Of course our photon mass
would be measured only when measuring the dispersion relation of
a polarized photon beam and would disappear when averaging over
polarizations. It is not an ordinary mass.

Another useful bound is the departure of the refraction index in
vacuum from unity, $i.e.$, $|\Delta n| =
|1-k/\omega|=|\gamma(\om)|/2$. In the 4D model, $|\Delta n^{(4D)}|
\simeq 4 |\Psi_0|$, In the 3D model, $|\Delta n^{(3D)}| \simeq
L_0/ 2\omega $ (when $L\ll L_0$). Generically Eq.~(\ref{e:gamma})
implies $|\Delta n| \leq 4.3 \times 10^{-31}  h_0
\nu^{-1}_{100}$. The difference of the refractive index from $1$
can be viewed as a difference of the photon speed from 1, $\Delta
c$ at the level of $10^{-30}$, which is much more stringent than
the (more general) limit  obtained in Ref. \cite{coleman}, which
is $\Delta c < 10^{-23}$. The formalism given here is applicable
also for higher dimension operators, but due to the frequency
dependence $|\alpha^{(LV)}| \propto \omega^{N-3}$, the CMB based
limits  on the amplitudes for higher dimension operators become
much weaker that those given from high energy photons ($\gamma$ or
$X$-rays). Even the bounds obtained from the nearby Crab Nebulae
are more promising \cite{crab} if $N\geq 5$.

\acknowledgments We appreciate useful comments and discussions
with A. Kosowsky, G. Lavrelashvili, and B. Ratra. R.D.
acknowledges support from the Swiss National Science Foundation.
T.K. acknowledges hospitality of Geneva University, partial
supports from GNSF grant ST06/4-096, and from the ICTP associate
membership program. R.D. and T.K. acknowledge the INTAS grant
061000017-9258. T.K. and Y.M. have received partial support from
DOE grant E-FG02-99ER41093.

\end{document}